\documentclass[]{aa}

\usepackage{natbib}
\usepackage{graphicx,xspace}
\usepackage{url}
\usepackage{threeparttable}
\usepackage{amsmath, amssymb}
\usepackage{color}
\usepackage{microtype}
\usepackage{booktabs, makecell}
\usepackage{subfig}
\usepackage{mwe}
\usepackage{lscape}
\usepackage[varg]{txfonts}
\usepackage{hyperref} 

\def\cxo{{Chandra}\xspace}
\def\ero{{eROSITA}\xspace}

\def\xmmn{{XMM-Newton}\xspace}

\newcommand\fergs{\ensuremath{\mathrm{erg}\,\mathrm{cm}^{-2}\,\mathrm{s}^{-1}}\xspace}

\newcommand\pspin{\ensuremath{P_{\rm spin}}\xspace}
\newcommand\porb{\ensuremath{P_{\rm orb}}\xspace}

\begin{document}

\title{PolarCat: Catalog of polars, low-accretion rate polars, and candidate objects}

\author{Axel D. Schwope}

\institute{Leibniz-Institut für Astrophysik Potsdam (AIP), An der Sternwarte 16, 14482 Potsdam, Germany \email{aschwope@aip.de}}
\date{Received 13 March 2025; accepted 23 April 2025}

\abstract{Polars and low-accretion rate polars (LARPs) are strongly magnetic cataclysmic variables. Mediated by the magnetic field of the white dwarf, their spin and binary orbit are (mostly) synchronized. They play an important role in our understanding of close binary evolution and the generation of strong magnetic fields in white dwarfs. Thanks to X-ray all-sky surveys, optical variability, and spectroscopic surveys, the number of polars and LARPs has grown from just a few in the 1980s to more than 200 today. Follow-up studies are facilitated by the systematic compilation of these systems presented here, which is also made available as an online resource. Yearly updates are planned, and community input is highly appreciated.}

\keywords{stars: cataclysmic variables -- stars: magnetic fields}

\maketitle

\section{Introduction }\label{s:intro} 

Polars are magnetic cataclysmic variables (MCVs). They harbor a magnetized white dwarf (WD), which accretes matter from a low-mass donor via Roche-lobe overflow (RLOF). The strong magnetic field keeps both stars in synchronous rotation; i.e., the spin period of the WD and the orbital period agree. It prevents the formation of an accretion disk and has a profound effect on the channels of energy release in the accretion regions via cyclotron cooling, which rivals thermal plasma radiation. Those semi-detached white dwarf/M dwarf (WDMD) binaries were nicknamed "polars" for their periodically variable strong optical polarization \citep{krzeminski_serkowski77}. For a general introduction on CVs and MCVs, see \cite{warner95}; and for an overview of the observational properties of polars, see \cite{cropper90}.

After their discovery in the late 1970s, several identified systems were studied in great detail to understand accretion physics in a strongly magnetic environment. For further details on the bright, early discoveries--such as the prototype AM Her, VV Pup (the soft X-ray machine), EF Eri (the textbook example), MR Ser, ST LMi, BL Hyi, QQ Vul, or V834 Cen--we refer to the hundreds of references available on NASA's Astrophysics Data Systems (ADS). Over the past one or two decades, the focus has shifted somewhat toward better understanding their collective properties and their role among the CVs.

As their distinct period distributions suggest \citep{schreiber+24}, polars evolve differently compared to non-MCVs. However, in the past, population synthesis has neglected the impact of magnetic fields \citep[see e.g.,][]{goliasch_nelson15}. Only recently, a first attempt to study MCV evolution using a revised angular momentum loss recipe was made by \cite{belloni+20}.

The fraction of MCVs among all CVs is uncertain but significant. The volume-limited sample by \cite{pala+20} shows that approximately one third of the systems are magnetic. On the other hand, the flux-limited \ero-based sample in the eFEDS field shows that approximately only 10\% of the systems are magnetic \citep{schwope+24}. However, both observed samples are limited by small total numbers and therefore may not be representative of the parent sample.

The origin of the magnetic field in CVs and pre-CVs is still under debate. Consensus nevertheless holds that the fields are not fossil, given the strong discrepancy between the large fraction of MCVs and the small fraction of detached magnetic WDMD pairs, as first brought to attention by \cite{liebert+05}. This implies that fields form at some point during CV evolution. Two scenarios were considered: the formation of magnetic fields during the common envelope phase \citep{tout+08}, or during the CV phase \citep{schreiber+21} itself.

In 1999, follow-up observations of peculiar objective prism spectra identified in the Hamburg-Quasar survey led to the discovery of a new subclass of MCVs. Their optical spectra were dominated by pronounced cyclotron harmonics \citep{reimers+99} in a field larger than that typically observed in standard polars. They were found to be X-ray emitters \citep[e.g.,][]{vogel+07}, yet their plasma temperatures and their mass accretion rates were considerably lower than those observed in "normal" polars. Initially regarded as normal polars at extremely low accretion rates, they were nicknamed LARPs, or Low-Accretion Rate Polars \citep{schwope+02b}. Later it became clear that at least some of the donor stars were Roche-lobe underfilling and that accretion was more likely mediated by a stellar wind instead of RLOF \citep{webbink_wickramasinghe05}, warranting their description as pre-polars \citep[PREPs,][]{schwope+09}. In this paper, these objects are designated simply as LARPs -- without implying any assumptions regarding their evolutionary status.

\begin{figure}[t]
\resizebox{\hsize}{!}{\includegraphics[clip=]{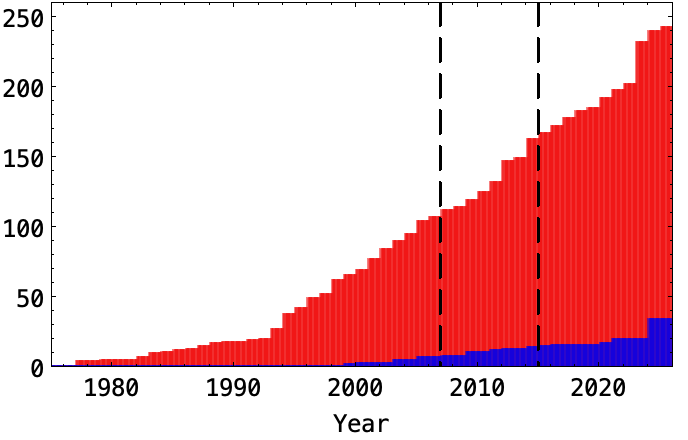}}
\caption{Growing population of polars and LARPs. The vertical dashed lines illustrate approximate dates of the final catalog compiled by \cite{ritter_kolb03} and the compilation of magnetic WDs by \cite{ferrario+15}. Here and in the following figures, red indicates polars, while blue indicates LARPs.
\label{f:ydisc}}
\end{figure}

Retaining an overview of known systems and their main class parameters is relevant for any systematic investigation. Examples concerning, for instance, their evolution were given above. The final 2007 edition of the well-known catalog compiled by \cite{ritter_kolb03} nevertheless lists 148 polars, including LARPs as well as some intermediate polars (IPs; see below). Around the same time, in 2006, the online catalog by \cite{downes+01} stopped compiling data. \cite{ferrario+15} presented a census of magnetic WDs--both single and in binaries. These authors listed 104 polars and ten LARPs and PREPs. Since then, the numbers have grown to more than 200 polars and more than 30 LARPs have been identified. The growth rate of both subclasses of MCVs is illustrated in Fig.\ref{f:ydisc}. The question of space densities and luminosity functions remains pressing, and the numbers derived are often uncertain by considerable factors \citep[see e.g.,][and references therein]{pala+20,pretorius+13,suleimanov+22}. Given the growing interest in population studies, an updated compilation in particular is both timely and necessary.

At the same time, we are witnessing a revolution in time-domain astronomy through all-sky surveys in the optical (CRTS, ZTF, ATLAS, ASAS-SN, GOTO, BlackGem, Gaia, SDSS, LAMOST; LSST, 4MOST, WEAVE in the future) and at X-ray wavelengths (\ero, \xmmn, \cxo), which deliver an ever-increasing number of new candidate or confirmed systems. Staying abreast of ongoing discoveries while retaining an overview requires reducing the amount of information per object to the essential minimum. This approach is adopted here in the hope that it will be useful to the community. The objects are listed here in two separate Tables available online: one listing objects that are either polars or candidate polars (i.e., semi-detached WDMD binaries with accretion through RLOF), and another listing LARPs (i.e., detached WDMD binaries with wind accretion). These Tables are described in the following sections. 

Intermediate polars are another important subclass of MCVs \citep[see e.g.,][and references therein]{hellier+93}. Similar to polars, they are semi-detached WDMD binaries with RLOF accretion. However, due to their longer orbital periods, they typically exhibit higher accretion rates and weaker magnetic fields; the spin of the WD is not synchronized with the binary orbital period. As these are catalogued on Koji Mukai's "The Intermediate Polars" homepage\footnote{\url{https://asd.gsfc.nasa.gov/Koji.Mukai/iphome/iphome.html}}, they are not considered here.

\section{The catalog of polars and polar candidates}

The list of polars (and candidate polars) was originally based on the catalog by \cite{ritter_kolb03}. Since then, it has been expanded through literature review as a private collection. An almost final, though preliminary, version of this list has been used in recent studies examining the orbital period distribution of MCVs and non-MCVs \citep{schreiber+24}, the periodic variability observed by the TESS satellite (Hernandez et al., subm.), and their appearance in the SRG/eROSITA X-ray all-sky survey \citep{schwope+24b}. Therefore, it seems appropriate to publish the catalog upon which such studies were based.

We must decide which objects to include in such a compilation and which properties or derived quantities are to be stored and made public. Although the definition of a polar is conceptually quite clear and simple, it is not so clear observationally. Defining properties include the seven following characteristics: (1) The orbital period lies between $\sim$80 min and $\sim$8 h; (2) The spin and orbital period are (almost) synchronized; (3) There is pronounced optical polarization originating from synchrotron radiation originating from the accreted matter; (4) X-rays originating from the cooling post-shock plasma show orbital phase-dependent modulation; (5) A strong soft X-ray source from the accretion area near the pole caps may be present; (6) High velocity emission lines of hydrogen as well as neutral and ionized helium appear with orbital phase-dependent line profiles; (7) Changes between states of high and low accretion may occur frequently at unpredictable intervals and with unpredictable duration.

All of these properties were identified in bright, early-discovered systems that were intensively studied across the entire electromagnetic spectrum, but rarely in newly discovered systems. In the context of surveys, classification is often based on a restricted set of data -- such as a single spectrum and a likely period, a typical light curve and a likely period, or a typical X-ray periodicity and spectrum -- which make it prone to uncertainties and revisions as more information is gathered. Such classifications nonetheless were accepted here, and thus polars and polar candidates were not distinguished. In a few cases, even a period could not yet be measured; however, a polar identification appeared highly likely or secure based on, for example, the observation of cyclotron harmonic humps or a typical emission line spectrum.

The catalog contains one line per object, and lists the following items per object:

\begin{itemize}
\item {\it Name:} The object's name as used by SIMBAD. If an object is not listed in SIMBAD, the name assigned by the discoverers is given.
\item {\it Alternative Name:} An alternative name of the object. This is not unique; for example, the prototype AM Her has more than 50 different identifiers. The alternative name may help find the object more easily. 
\item {\it Gaia DR3:} The source ID from the third Gaia catalog. Newer detections are often fainter than the Gaia limit and therefore lack this information.
\item {\it RA \& Dec:} The system's coordinates (right ascension, declination) in the equatorial system (FK5, ep=J2000, eq=2000).
\item {\it $P_{\rm orb}$:} The period listed is assumed to be the orbital period, although in many cases this is not firmly established. Unique determination of the orbital period is possible for eclipsing systems via time-resolved photometry or from radial velocity curves tracing a unique feature originating from the donor star, usually the Na doublet in the near IR or the narrow emission line from the irradiated front side. The number of digits reported reflects the accuracy given in the source under {\it Ref\_{\rm period}}. 
If only a single period is detected from time-resolved photometry of a non-eclipsing system, it is taken to represent the spin period, with spin-orbit synchronism typically assumed. Systems known to be slightly asynchronous are marked "AP" in the comment column. The assumed degree of asynchronism is generally low, $|\frac{\pspin - \porb}{\pspin}| < 0.05$. An object such as EX Hya, with $P_{\rm spin}/P_{\rm orb} = 2/3$, is considered an IP, while more typical IPs have $P_{\rm spin}/P_{\rm orb} = 0.1$. However, a growing subclass of EX Hya-like objects have a significantly higher period ratio $P_{\rm spin}/P_{\rm orb}$  above the canonical value of 0.1. This blurs the clear distinction between IPs and polars \citep[see e.g.,][]{pradeep+24,littlefield+23}. 

\begin{figure}[t]
\resizebox{\hsize}{!}{\includegraphics[clip=]{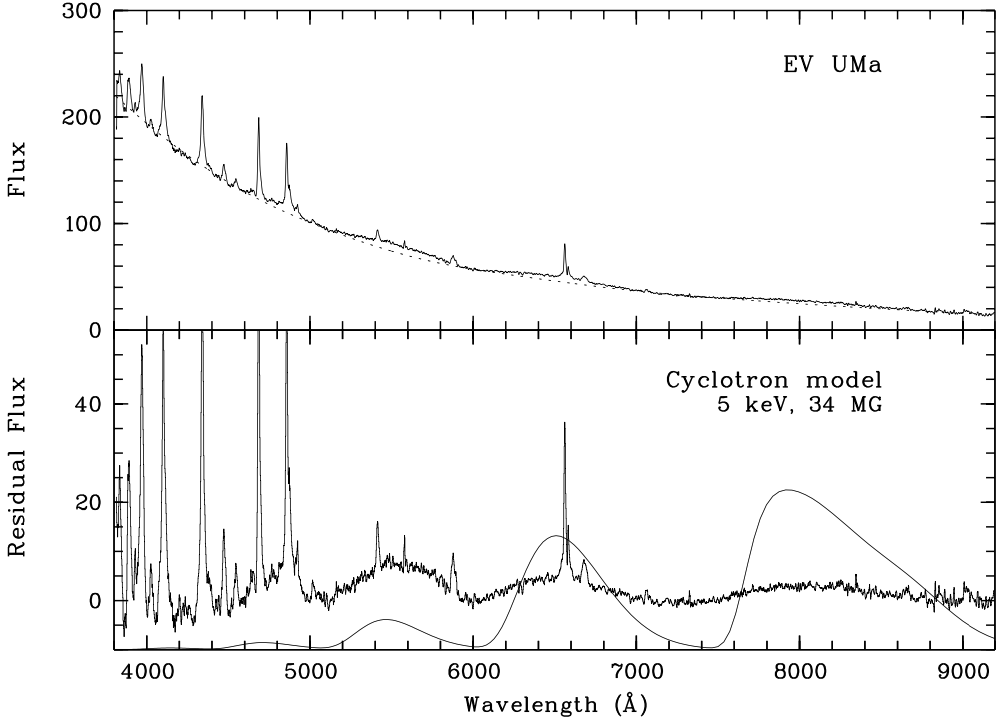}}
\caption{SDSS spectrum of EV UMa showing cyclotron harmonics. The upper panel shows the original spectrum (solid line) and a low-order polynomial fit to the continuum (dashed line). Flux units are $10^{-17}$\,\fergs. The lower panel shows the difference spectrum between the original and the continuum fit, together with an arbitrarily scaled 5 keV cyclotron model. The model was shifted by ten flux units. 
\label{f:evuma}}
\end{figure}

\item {\it $B_1, B_2, B_{\rm halo}, B_{\rm phot}$:} The magnetic field is the second essential parameter of a polar. It can be measured either through cyclotron resonances (harmonics) observed in optical, IR, or UV spectra, or via Zeeman-shifted Balmer lines. An example of cyclotron harmonics found in the SDSS spectrum of EV UMa is given in Fig.~\ref{f:evuma}. 
The continuum of EV UMa was found to be modulated by low-amplitude and low-frequency undulations. These became much more pronounced after subtraction of a smooth continuum and could be identified with cyclotron harmonics in a field of 34 MG. The model shown in the lower panel of the figure is for illustrative purposes only, but it reproduces the observed wavelengths of the cyclotron harmonics very well. An estimate of the field strength in the range of 30-40 MG was already provided in the discovery paper by \cite{osborne+94}. Thanks to the wide wavelength coverage and the high signal-to-noise ratio of the SDSS spectrum shown in Fig.~\ref{f:evuma}, the identification of the cyclotron harmonics and the value of the magnetic field strength now appear unique.

While the cyclotron lines originate from the cooling accretion plasma, the Zeeman lines may originate either from a cool halo surrounding the accretion column or region, or from the WD photosphere. Polars may accrete onto two poles (polar regions), and some polars exhibit cyclotron lines from each of the two accretion columns. The derived field strengths are listed under the columns labeled $B_{\rm 1}, B_2$. The measured field of the Zeeman halo lines is listed under the $B_{\rm halo}$ column, and the mean field measured from the photosphere is listed under $B_{\rm phot}$. The magnetic field structure on the WD surface is often non-dipolar, and the implied polar field strength can differ from the reported mean field \citep[see e.g.,][]{beuermann+07}.

\item {\it ecl\_len:} If the object shows binary star eclipses (e.g., a WD obscured by the donor star), the eclipse length is given in minutes. This value is often only an estimate and should therefore be used with caution. For comparative purposes, it would be ideal to provide the FWHM of the eclipse of the WD. However, this is not often measured or not measurable. Instead, the width of the eclipse of the accretion spot on the WD is observable and differs from the eclipse length of the WD. If the eclipse length cannot be obtained from the corresponding papers, it is estimated from published light curves. A value of 99 indicates that an eclipse was reported, but no further information about its length is provided.

\item {\it Ref\_disc:} Reference made to a discovery paper. The bibcode provided is a unique identifier assigned by ADS to each paper. It references the first paper mentioning a detected source as most likely or definitely belonging to the polar class, but this is not in any case unique.
\item {\it Ref\_period:} The bibcode of the paper reporting the given period (Col. \#6).
\item {\it Ref\_B:} The bibcode of the paper or papers reporting the magnetic field strengths in Cols. 7-10.  
\item {\it Comment:} A brief comment about the source, or some other relevant short note. AP stands for asynchronous polar, to be understood as asynchronous MCV.
\end{itemize}

\section{The catalog of LARPs}

While ordinary polars were found in various ways, LARPs were discovered in two ways, either through their extraordinary optical spectra or, more recently \citep{vanroestel+25}, through their pronounced optical light curves, with both effects being a direct consequence of cyclotron radiation in the accretion plasma. Whether an object that was once identified as a LARP remains a LARP is uncertain, once more data are collected over a longer period of time base. As an example, if EF Eri were first observed between 1997 and 2022 in its extended low state, it would not be recognized as an ordinary polar, but perhaps as a LARP. The only reliable way to classify with certainty an object as a LARP -- a semi-detached WDMD binary -- is to measure the radius of the donor star, which must be Roche-lobe underfilling. This measurement is difficult to perform with accurate precision. A classification as a LARP might, therefore, be considered preliminary. The object might be an ordinary polar in an extended low state. However, objects that are currently considered LARPs are listed in the second catalog. 

The catalog has one line per object, and lists the following items per object:

\begin{itemize}
\item {\it Name:} The name of the object as used by SIMBAD. If an object is not listed in SIMBAD, it is the name given by the discoverers.
\item {\it Alternative Name:} An alternative name of the object. 
\item {\it Gaia DR3:} The source ID in the third Gaia catalog. 
\item {\it RA \& Dec:} Coordinates (right ascension, declination) of the system in the equatorial system (FK5, ep=J2000, eq=2000) 
\item {\it $P_{\rm orb}$:} The period given is assumed to be the orbital period, although in many cases (as for the polars) there is a lack of firm proof. The number of digits reported corresponds to that given in the {\it Ref\_{period}} reference.
\item {\it $B_1, B_2, B_{\rm Zeeman}$:} Most magnetic fields in LARPs are measured with the help of cyclotron resonances (harmonics) in the optical wavelength range. LARPs may have one or two accretion regions, and cyclotron lines are observed in some objects from both regions (e.g., magnetic field strength listed in the $B_{1,2}$ columns). There is one pre-CV candidate with Zeeman split emission lines, and two additional pre-CV candidates with Zeeman-split photospheric absorption lines (field strength listed in the $B_{\rm Zeeman}$ column).  
\item {\it Ref\_disc:} Reference to the discovery paper (bibcode assigned by ADS). 
\item {\it Ref\_period:} The bibcode of the paper reporting the given period.
\item {\it Ref\_B:} The bibcode of the paper reporting the magnetic field strengths.  
\item {\it Comment:} A brief comment on the source, or some other relevant short note.
\end{itemize}

\section{Results}
The catalogs published here list 242 polars and 33 LARPs. In this section, we first discuss some results on polars.

\begin{figure}[t]
\resizebox{\hsize}{!}{\includegraphics[clip=]{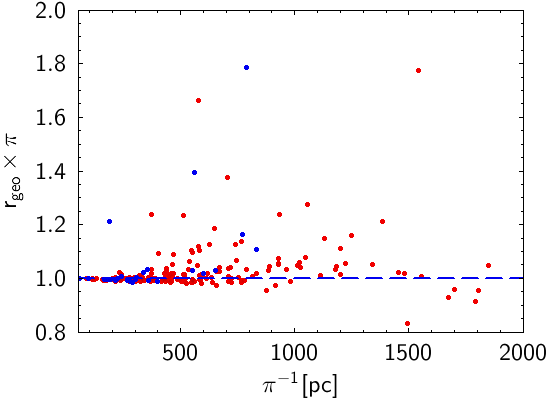}}
\caption{Distance estimates of polars. Shown here along the $y$-axis is the ratio $r_{\rm geo}$ \citep{bailer-jones+21} over the inverted parallax as a function of the inverted parallax for polars with $\pi^{-1} < 2$ kpc.
\label{f:dist}}
\end{figure}

The growing number of discoveries is illustrated in Fig.~\ref{f:ydisc}. Between 1995 and 2022, the average discovery rate for polars was about six. There was a boost in 2023, when more than 30 were reported for the first time by \cite{inight+23a} and \cite{inight+23b}. This was only an apparent burst of new discoveries and was the result of a comprehensive assessment of all SDSS CV spectra obtained over the past more than twenty years. These were not contained in Szkody's final list from 2011 \citep{szkody+11}. 

\begin{figure}[t]
\resizebox{\hsize}{!}{\includegraphics[clip=]{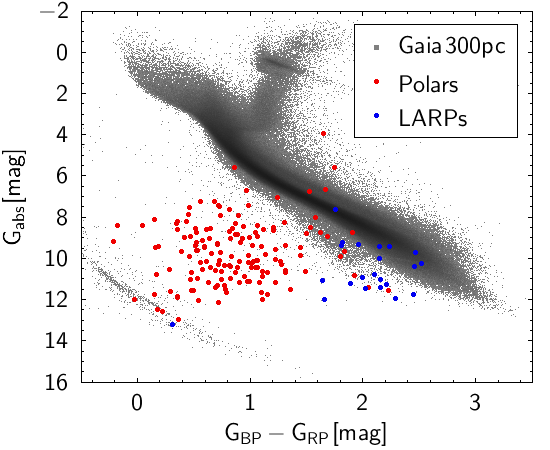}}
\caption{Gaia-based color-magnitude diagram of polars (red symbols) and LARPs (blue symbols). Background objects consist of unrelated objects drawn from the Gaia catalog within 300 pc, included for reference. Only objects with distance estimates agreeing to within 10\% are shown. No extinction correction is applied.}
\label{f:cmd}
\end{figure}

Only about 10\% of the polars do not have a Gaia DR3 ID, which means that for most polars a distance estimate can be given. In Fig.~\ref{f:dist}, two distance estimates are compared for the 184 polars with $\pi^{-1} < 2$\,kpc: the distance $r_{\rm geo}$ derived by \cite[][BJ21]{bailer-jones+21} and the inverted parallax. In all the figures of this paper, red indicates polars, and blue indicates \begin{table}[ht]
\caption{Distance estimates for polars.}
\label{t:dist}\centering
\begin{tabular}{c|c|l|l|l}
\hline\hline
Distance range [pc] & No & Mean & StDev & Median\\
\hline
$<350$ & 45 & 0.997 & 0.007 & 0.996\\
350 -- 500 & 35 & 1.016 & 0.046 & 0.999 \\
500 -- 1000 & 63 & 1.05 & 0.11 & 1.013\\
1000 -- 2000 & 26 & 1.06 & 0.18 & 1.04\\
\hline
\end{tabular}
\tablefoot{
 Given are the mean, the standard deviation, and the median of the ratio $r_{\rm geo} * \pi$ for the specified number of polars within four distance shells. EU Cnc was excluded from the 2nd shell.
}
\end{table}
LARPs. Sample properties (e.g., mean, standard deviation, and median) of the ratio of the two distance estimates are given in Table \ref{t:dist} for four adjacent distance shells. Disagreements between the two methods used to derive or infer a distance may be due to low data quality, the binary nature of the sources -- which can impact astrometric parameters and result in nonstandard locations in the color-magnitude diagram -- and, thus, inappropriate priors when inferring distances from parallaxes, or from a combination of these effects. The nearest polar is AM Her at 87.9 pc. For objects with $\pi^{-1} < 350$\,pc (45 objects), both distance estimates agree within a few percent. The distances by BJ21 are typically shorter than inverted parallaxes. This mostly also holds true between 350 pc and 500 pc; both methods typically provide a good distance estimate. Only five of 35 estimates in this distance range differ by more than 5\%. There is one marked outlier, EU Cnc, that has $\pi^{-1} = 408$\,pc and $r_{\rm geo} = 2360$\,pc. However, its {\tt parallax\_over\_error} is only 1.49, which is a very uncertain measurement. The figure suggests that polar distances can be determined with reasonable accuracy for individual objects up to 1 kpc, but beyond that distance, accuracy can only be achieved through a sample average.

Periods are listed for 215 out of 242 polars and range from 77.8 min to 855 min. Twenty-four polars (10\%) exhibit a period below the CV binary orbit period minimum (i.e., below 82.4 min) determined by \cite{knigge+11}. This indicates a different angular momentum loss rate acting in MCVs compared to non-MCVs. The catalog contains one object with a period longer than that of V1309 Ori (478 min): V479 And with a period of 855 min. While the nature of V1309 Ori as a polar is undisputed -- despite having a nonstandard donor -- the nature of V479 And is not yet clear. \cite{gonzalez-buitrago+13} discuss it as a possible wind-accreting system. Hence, it could also be classified as a LARP, but the optical spectra that show strong He{\sc II} emission lines suggest that it is likewise a nonstandard LARP.

Eleven or 12 systems ($\sim$5\%) are known APs. This percentage is considered a lower limit because independent spin and orbit measurements are not available for all objects. Only one object is reported to have a spin period longer than the orbital period, V1432 Aql. The degree of asynchronism varies from $\pspin/\porb = 0.8028$ (Swift J0503.7-2819) to 1.0028 (V1432 Aql). Two APs are found at an orbital period of 81 min, with the longest-period AP at 277 min. Thus, APs are not preferentially found at either long or short periods, which otherwise could suggest a reason for their asynchronism. Five APs have a measured magnetic field from cyclotron harmonics in the range of 11 and 71 MG, similar to the values found in synchronous polars.

About 15\% of polars (35 objects) are eclipsing, a fraction consistent with expectations for randomly oriented semi-detached CVs. The measured eclipse length varies between 3 min to 40 min. Objects with shorter eclipse length seem to be missing due to their faintness and the long exposure times used for obtaining light curves. By consequence, shorter eclipses were simply overlooked. The relative eclipse length, $\Delta t_{\rm ecl} / P_{\rm orb}$, varies between 0.03 and 0.09. 

\begin{figure}[t]
\resizebox{\hsize}{!}{\includegraphics[clip=]{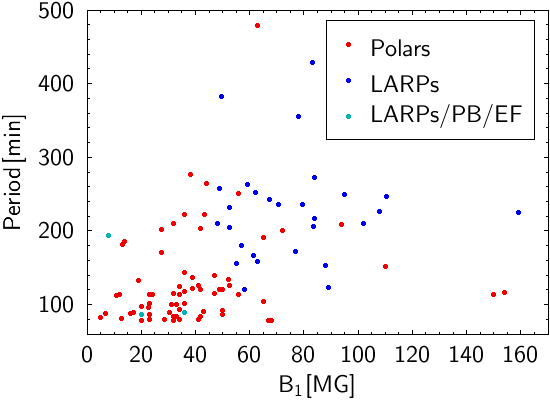}}
\caption{Magnetic field $B_1$ at the main accretion spot of the polars (red symbols) and the LARPs (blue symbols) as a function of the orbital period. For LARPs, the magnetic field was determined through cyclotron harmonics. The three objects with cyan symbols are LARP candidates whose fields strengths were measured through Zeeman split Balmer lines.
\label{f:b1vsp}}
\end{figure}

We now turn to a short discussion of the LARPs. They first appeared in 1999 (see Fig.~\ref{f:ydisc}), and their growth rate was moderate. New detections were made mainly through ongoing SDSS spectroscopic surveys. Only recently did ZTF deliver a comparatively large number of new discoveries \citep{vanroestel+25}. All LARPs have a Gaia DR3 ID. Among the current sample of (candidate) LARPs and polars, ATO J026.6489+49.2452 is the closest object. It is found at an astonishingly short distance of 56.6\,pc, a recent discovery identified as an apparently variable WD \citep{guidry+21}. Within 550 pc, inverse parallaxes and BJ21 distances agree to within a few percent, with two exceptions: 2CXO J182117.2-131405 and SDSS J013714.97+210220.0, which have $\pi^{-1} = 190$\,pc and 356 pc, but $r_{\rm geo} =229$\,pc and 793 pc, respectively. Beyond 550 pc, distance estimates become rather uncertain (see Fig.~\ref{f:dist}).

The color-magnitude diagram shown in Fig.\ref{f:cmd} was constructed using only systems for which the distances derived from inverse parallaxes and from BJ21 agree to better than 10\%. LARPs and polars appear rather well separated in the CMD. Most polars are located between the WD and the main sequence with a few marked outliers. Some polars are found on the WD sequence. These are polars that were in low states during the complete Gaia survey. The brightest object--discussed above--is V479 at $G=5.5$. By its extraordinarily long period, it appears to be a borderline object. The next brightest object at $G=6.6$ is V1309 Ori. All LARPs have very red colors, $B-R> 1.5$, and are well separated from the much bluer ordinary polars. The one exception is the closest MCV, ATO J026.6489+49.2452, which lies directly on the WD sequence and was indeed identified as a variable WD. 

All but two LARPs have long periods ($P>120$\,min). The two short-period objects, SDSS J125044.42+154957.3 at 86.3 min and SDSS J151415.65+074446.5 at 88.7 min, also have low field strengths at $B=20$ and 36 MG, respectively. The field strength was measured through photospheric Balmer absorption lines split via the Zeeman effect. They appear distinct from other LARPs in a $(P,B)$ diagram (Fig.~\ref{f:b1vsp}). Whether these objects truly represent an extension of the class toward the lower left in that figure, or if they need to be classified as something else remains open. They could be ordinary polars in an extended low state as EF Eri, or they could be period bouncers \citep[PBs; for a discussion of these objects, see][]{breedt+12}. A third object with low field strength, SDSS J030308.35+005444.1, also appears distinct from the more typical high-field LARPs \citep{parsons+13}. Its field of only 8 MG was measured through Zeeman-split Balmer emission lines originating from gaseous matter confined between the M5 companion and the magnetic WD. This measurement likely represents a lower limit to the field at the WD surface and is not directly comparable to those in systems where fields were inferred from cyclotron resonances.
\begin{figure}[t]
\resizebox{\hsize}{!}{\includegraphics[clip=]{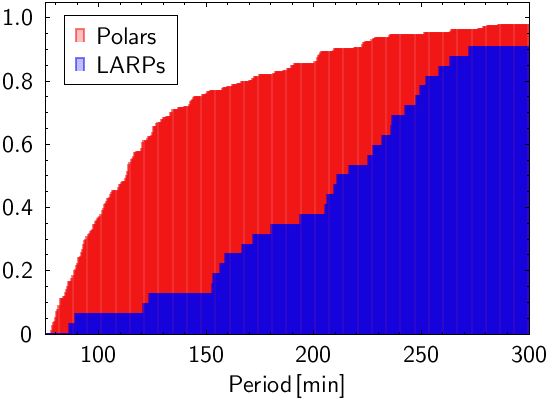}}
\caption{Cumulative, normalized period distributions of polars and LARPs
\label{f:pcum}}
\end{figure}

The locations of polars and LARPs in a ($B,P$) scatter diagram and their observed cumulative normalized period distributions are strikingly different (Figs.~\ref{f:b1vsp} and \ref{f:pcum}). The three low-field and short-period LARPs discussed in the previous paragraph, which do not seem to completely comply with the other LARPs, are marked with a different color in Fig.~\ref{f:b1vsp}. They are labeled as LARPs/PB/EF to indicate that they may be LARPs, period bouncers, or EF Eridani-like ordinary polars in extended low states. Polars are discovered mainly at short periods (65\% below 126 min), while LARPs are observed mainly at longer periods. LARPs have high fields, while polars have lower fields (65\% of the polars have a field strength below 42 MG). Care must be taken not to overinterpret observed distributions that are influenced by observational biases. The lack of low-field LARPs is due to the shift of the cyclotron spectrum into the infrared at low field strengths \citep{schwope+09}. Such systems are therefore missing in such optical photometric and spectroscopic surveys as the ZTF and the SDSS, which otherwise provide most known LARPs. The lack of short-period LARPs is not easily explained by an observational bias and could suggest that such systems have not yet evolved to short periods. The only alternative seems to be that the donor star's wind -- and hence observational features -- might be so weak that these systems have simply been overlooked so far. Polars do not show an obvious period gap, but we still lack a sufficient number of long-period polars in a volume-limited sample to make a definitive statement. The period distribution of the LARPs is not sufficiently well populated to allow even preliminary conclusions about the presence or absence of certain features.

\section{Discussion and conclusion}
This paper provides up-to-date lists of polars, LARPs, and candidate objects, totaling 242 polars and 33 LARPs. While, inevitably, the compilation will be incomplete already upon publication, sharing this data serves the community. Updates will be planned annually and community feedback is encouraged. Depending on the volume of suggested revisions, the first update may appear much sooner.

All values reported in the Tables reflect the best-available knowledge at the time of writing. Nevertheless, some relevant studies, revisions to older periods, and determinations of the magnetic field may have been missed. Any suggestions or corrections are welcome, and I hope that these lists will be useful.

\section{Data availability}
The two catalogs are available to the community as VO Tables on the author's homepage, \url{https://www.aip.de/de/members/axel-schwope/polarcat}.

\begin{acknowledgements} 
Comments given by an anonymous referee helped improve the quality of the article and are gratefully acknowledged.

The help of students and undergraduates
Thijs Dorhout,
Fabian Emmerich, 
Kira Knauff,
Timon Thomas (all AIP),
Daniela Munoz-Giraldo, and
Sebastian Hernandez (IAAT Tübingen) 
in compiling the data presented here is gratefully acknowledged. 

The author acknowledges the partial support of the Deutsches Zentrum für Luft- und Raumfahrt (DLR) under Contract Nos.~50 OX 1401, 50 QR 1604, 50 OX 1901,50 QR 2104, 50 OX 2203, and 50 OX 2301. 
\end{acknowledgements} 

\bibliographystyle{aa}
\bibliography{cat_polars}%
\end{document}